\documentclass[a4paper]{jpconf}
\usepackage{graphicx}

\newcommand{\apj}{{ApJ}}			
		
\newcommand{\apjs}{{ApJS}}		
		
\newcommand{\aap}{{A\&A.}}

\newcommand{\prc}{{Phys.~Rev.~C}}

\newcommand{\pasp}{{PASP}}

\newcommand{\physrep}{{Phys.~Rep.}}

\begin{document}
\title{Nuclear astrophysics with radioactive ions at FAIR}

\author{
R~Reifarth$^{1}$,
S~Altstadt$^{1,2}$,
K~G\"{o}bel$^{1}$,
T~Heftrich$^{1,2}$,
M~Heil$^{2}$,
A~Koloczek$^{1,2}$,
C~Langer$^{1,2,60}$,
R~Plag$^{1,2}$,
M~Pohl$^{1}$,
K~Sonnabend$^{1}$
% ...
}
\address{$^{1}$~University of Frankfurt, Germany}
\address{$^{2}$~GSI Darmstadt, Germany}
% ...
\address{$^{60}$~NSCL/MSU, USA}
% ...
\ead{reifarth@physik.uni-frankurt.de}

\begin{abstract}
The nucleosynthesis of elements beyond iron is dominated by neutron captures in the $s$ and $r$ processes. However, 32 stable, proton-rich isotopes cannot be formed during those processes, because they are shielded from the $s$-process flow and $r$-process $\beta$-decay chains. These nuclei are attributed to the $p$ and $rp$ process. 

For all those processes, current research in nuclear astrophysics addresses the need for more precise reaction data involving radioactive isotopes. Depending on the particular reaction, direct or inverse kinematics, forward or time-reversed direction are investigated to determine or at least to constrain the desired reaction cross sections.

The Facility for Antiproton and Ion Research (FAIR) will offer unique, unprecedented opportunities to investigate many of the important reactions. The high yield of radioactive isotopes, even far away from the valley of stability, allows the investigation of isotopes involved in processes as exotic as the $r$ or $rp$ processes. 
\end{abstract}

\section{Introduction}
Radioactive beams offer the opportunity to extend the experimentally based knowledge about nuclear structure
far beyond the valley of stability. Especially within the planned international 
Facility for Antiproton and Ion Research (FAIR) at GSI \cite{FAIR}, radioactive ions
will be produced with highest intensities. It is very often not feasible to collect the respective radioactive 
ions in order to produce a sample for irradiation with e.g. neutrons, protons or gammas. 
Experiments in inverse kinematics - irradiating a stable target with the desired radioactive ions - are the solution
to that problem. In this article, we will mostly report about performed and upcoming in-beam experiments and not about the wide field of possible ring experiments.

The proposed R$^3$B setup \cite{R3B}, a universal setup for kinematically complete measurements of
Reactions with Relativistic Radioactive Beams 
will cover experimental reaction studies with exotic nuclei far off stability, 
with emphasis on nuclear
structure and dynamics. Astrophysical aspects and technical applications are also concerned.
R$^3$B is a versatile reaction setup with high efficiency, acceptance, and resolution for reactions with high-energy radioactive beams. The setup will be located at the High Energy Cave which follows the 
high-energy branch of the new fragment separator (Super-FRS). The experimental configuration is based on
a concept similar to the existing LAND setup at GSI introducing substantial improvement with
respect to resolution and an extended detection scheme, which comprises the additional detection efficiency of light
(target-like) recoil particles and a high-resolution fragment spectrometer. The setup is adapted to the
highest beam energies (corresponding to 20 Tm magnetic rigidity) provided by the Super-FRS
capitalizing on the highest possible transmission of secondary beams. The experimental setup is suitable
for a wide variety of scattering experiments, such as heavy-ion induced electromagnetic excitation,
knockout and breakup reactions, or light-ion (in)elastic and quasi-free scattering in inverse kinematics,
thus enabling a broad physics program with rare-isotope beams to be performed \cite{R3B}.

Applying the Coulomb dissociation method \cite{BaR96,BeG10} R$^3$B contributes already now to almost every astrophysical scenario. With the expected increase in the production of radioactive species at FAIR, even more exotic reactions can be investigated. Both, the current situation and the prospects at FAIR are shown in section~\ref{section_60fe} for the $s$ process,  in section~\ref{section_b} for the $r$ process and in section~\ref{section_30s} for the rp process. Recent experiments and future prospects investigating charge exchange reactions are discussed in section~\ref{section_152eu} for the $s$ process and section~\ref{section_64ge} for the $\nu$p process.

\section{$^{60}$Fe - a product of the $s$ process}\label{section_60fe}
A significant contribution to
the interstellar abundance of the radiogenic $^{60}$Fe  is provided by the slow
neutron capture (s) process in massive stars, the weak component of the $s$ process. The s
process in massive stars operates in two major evolutionary stages, first
during convective core He-burning and, subsequently,
during convective shell C-burning. Neutrons are mainly
produced by the $^{22}$Ne($\alpha$,n) reaction in both cases,
but at rather different temperatures and neutron densities
\cite{PGH10,KGB11}.

\begin{figure}
\begin{center}
  \includegraphics[width=.995\textwidth]{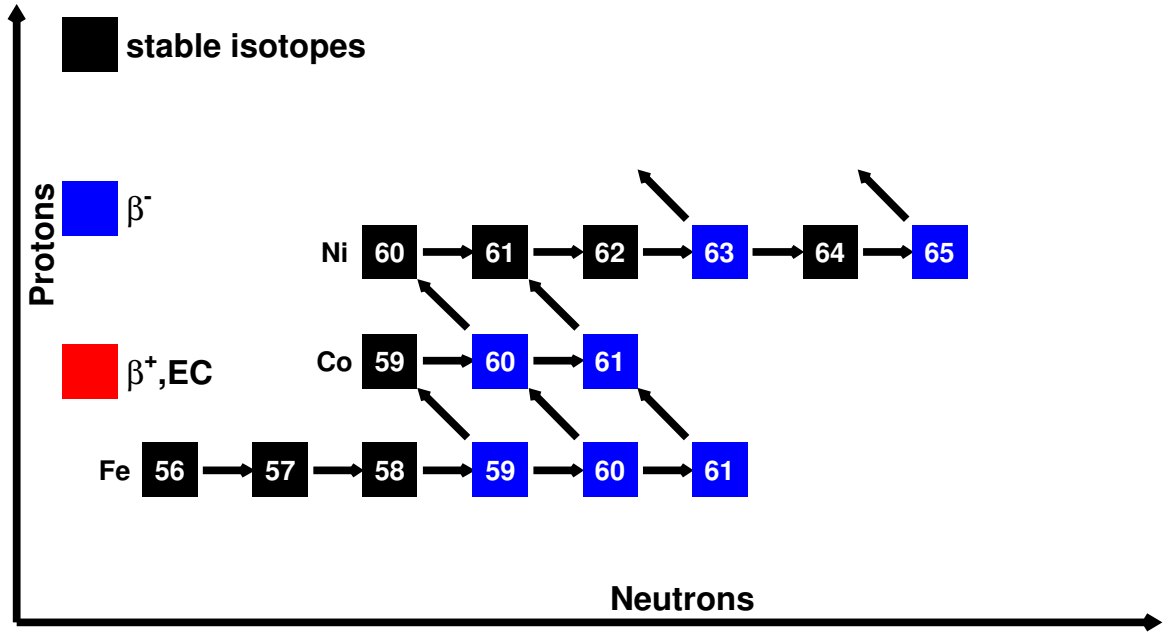}
\end{center}
  \caption{The $s$-process path between Fe and Co. The neutron densities during the $s$ process have to be sufficiently high to overcome the rather short-lived isotope $^{59}$Fe ($t_{1/2}=45$~d). 
  \label{s_process_fe-co}}
\end{figure}

As illustrated in Figure~\ref{s_process_fe-co}, the $s$-process path to $^{60}$Fe,
which starts from the most abundant seed nucleus $^{56}$Fe, is
determined by the branching at $^{59}$Fe (t$_{1/2}$~=~44.5~d). At
the low neutron densities during convective core He burning,
$^{60}$Fe is shielded from the $s$-process chain, because the
$\beta^-$-decay rate of $^{59}$Fe dominates over the (n,$\gamma$) rate by orders
of magnitude. On the other hand, the production of
$^{60}$Fe becomes efficient during the shell C-burning phase,
where higher temperatures of T~=~$(1.0-1.4)\cdot 10^9$~K
give rise to the neutron densities in excess of $10^{11}~$cm$^{-3}$
necessary for bridging the instability gap at $^{59}$Fe.
The interpretation of all the above observations depends
critically on the reliability of the stellar models as
well as on the reaction rates for neutron capture relevant
to the production and depletion of $^{60}$Fe \cite{DPB06}. These
rates can only be determined reliably in laboratory experiments,
because theoretical calculations are too uncertain. 
Since the neutron capture  cross section of $^{60}$Fe(n,$\gamma$) has been measured in the astrophysically interesting energy region \cite{URS09} and a reliable value for the half life of the $\beta^-$-decay has been provided \cite{RFK09}, the most important missing piece to understand the stellar production of $^{60}$Fe is the $^{59}$Fe(n,$\gamma$) cross section under stellar conditions, see Figure~\ref{s_process_fe-co}.

\begin{figure}
\begin{center}
  \includegraphics[width=.995\textwidth]{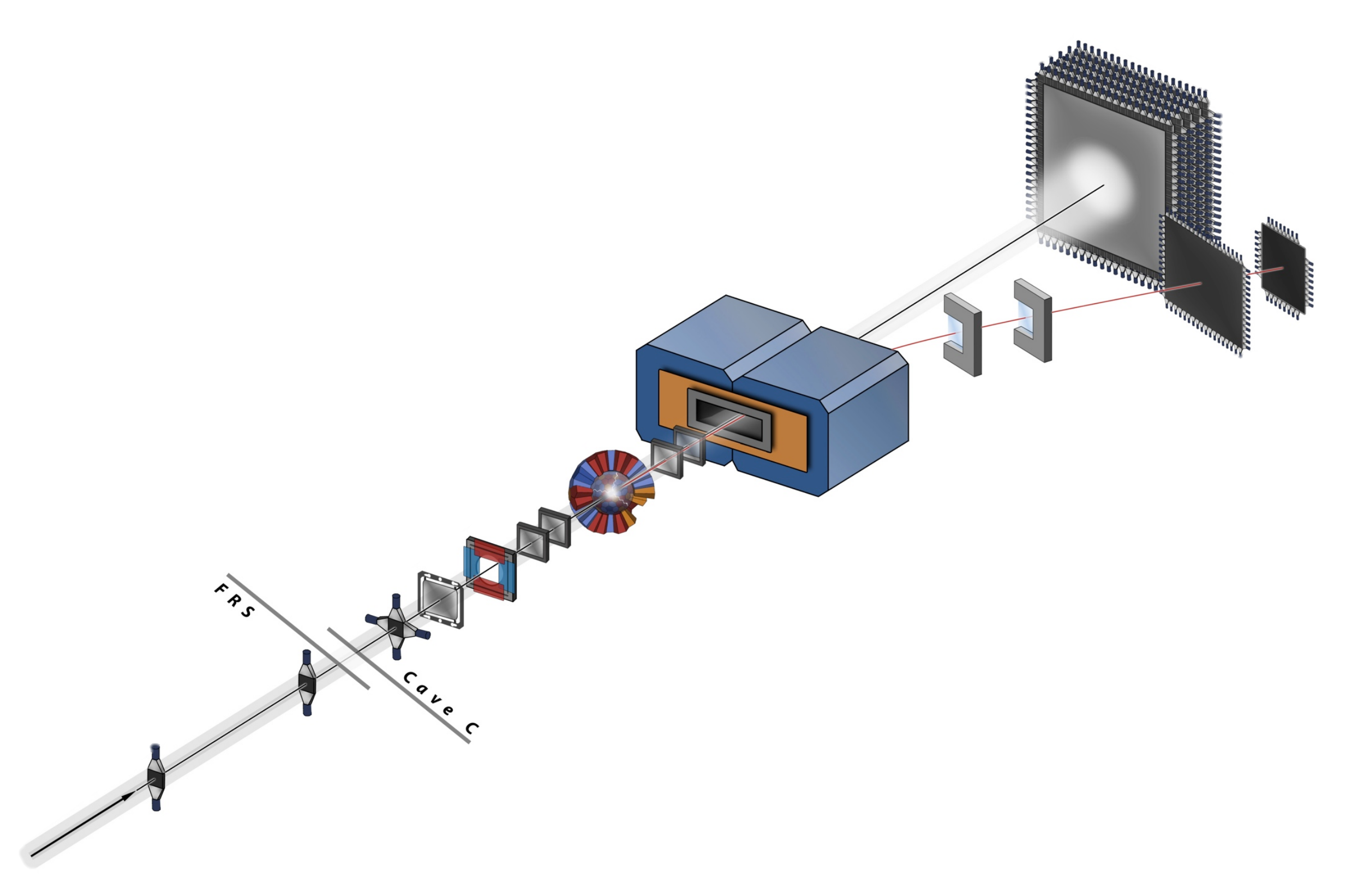}
\end{center}
  \caption{The R$^3$B setup at GSI optimized for detecting neutrons in the exit channel. The beam enters Cave C after passing the FRS (bottom, left), passes several detectors used for incoming identification and hits the target in the center of th crystal ball (half of the ball is drawn). Afterwards charged fragment and neutrons are separated by the ALADIN magnet. The fragments are bent to the right analysed with a suite of scintillator detectors while the neutrons remain unchanged and are detected with LAND.
  \label{r3b_setup_heavyFragments}}
\end{figure}

Because the half-life of $^{59}$Fe is only 45~d, indirect methods have to be applied to determine the neutron capture cross section. Since $^{60}$Fe is unstable too, the method of choice is the determination of the desired A(n,$\gamma$)B via the inverse reaction B($\gamma$,n)A applying the Coulomb dissociation (CD) method at the LAND/R$^3$B setup (Figure~\ref{r3b_setup_heavyFragments}).

\begin{figure}
\begin{center}
  \includegraphics[width=.995\textwidth]{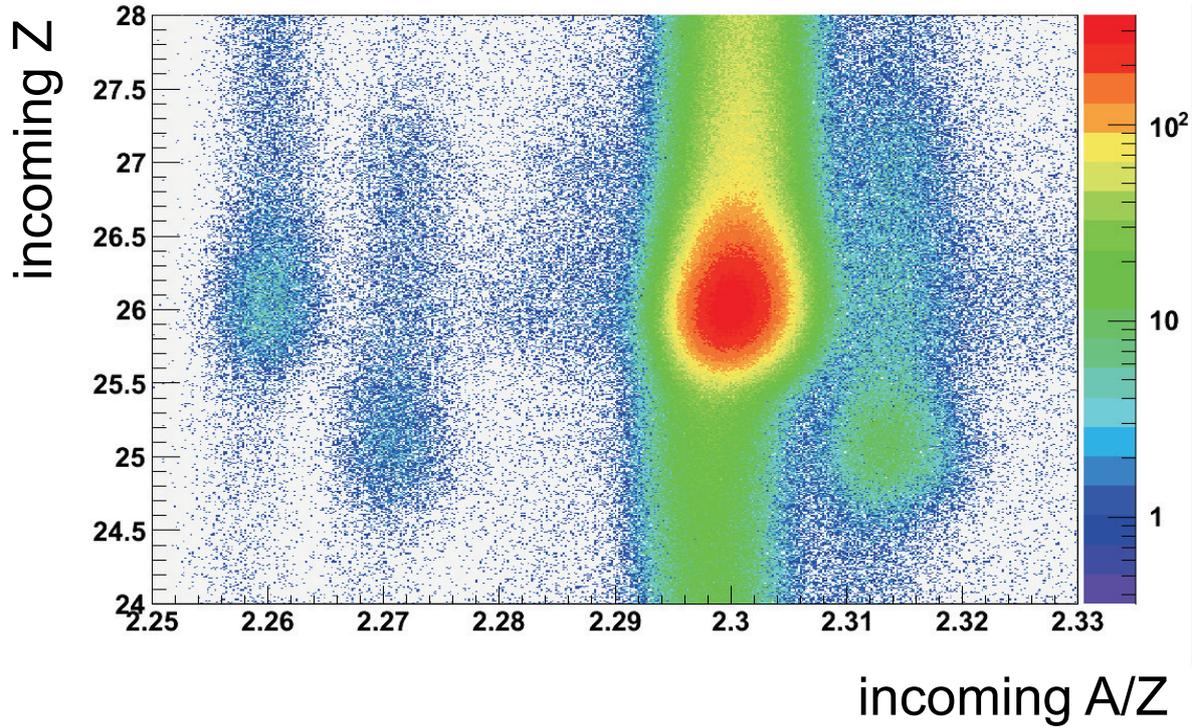}
\end{center}
  \caption{Incoming particle identification at the R$^3$B setup. 
  \label{incoming_60fe}}
\end{figure}

$^{60}$Fe ions have been produced by fragmentation of $^{64}$Ni. After passing the fragment separator (FRS, \cite{GAB92}) most of the unwanted species are removed and a beam consisting of almost only $^{60}$Fe arrives in at the LAND/R$^3$B setup, where each ion is identified in charge and mass, Figure~\ref{incoming_60fe}.

\begin{figure}
\begin{center}
  \includegraphics[width=.475\textwidth]{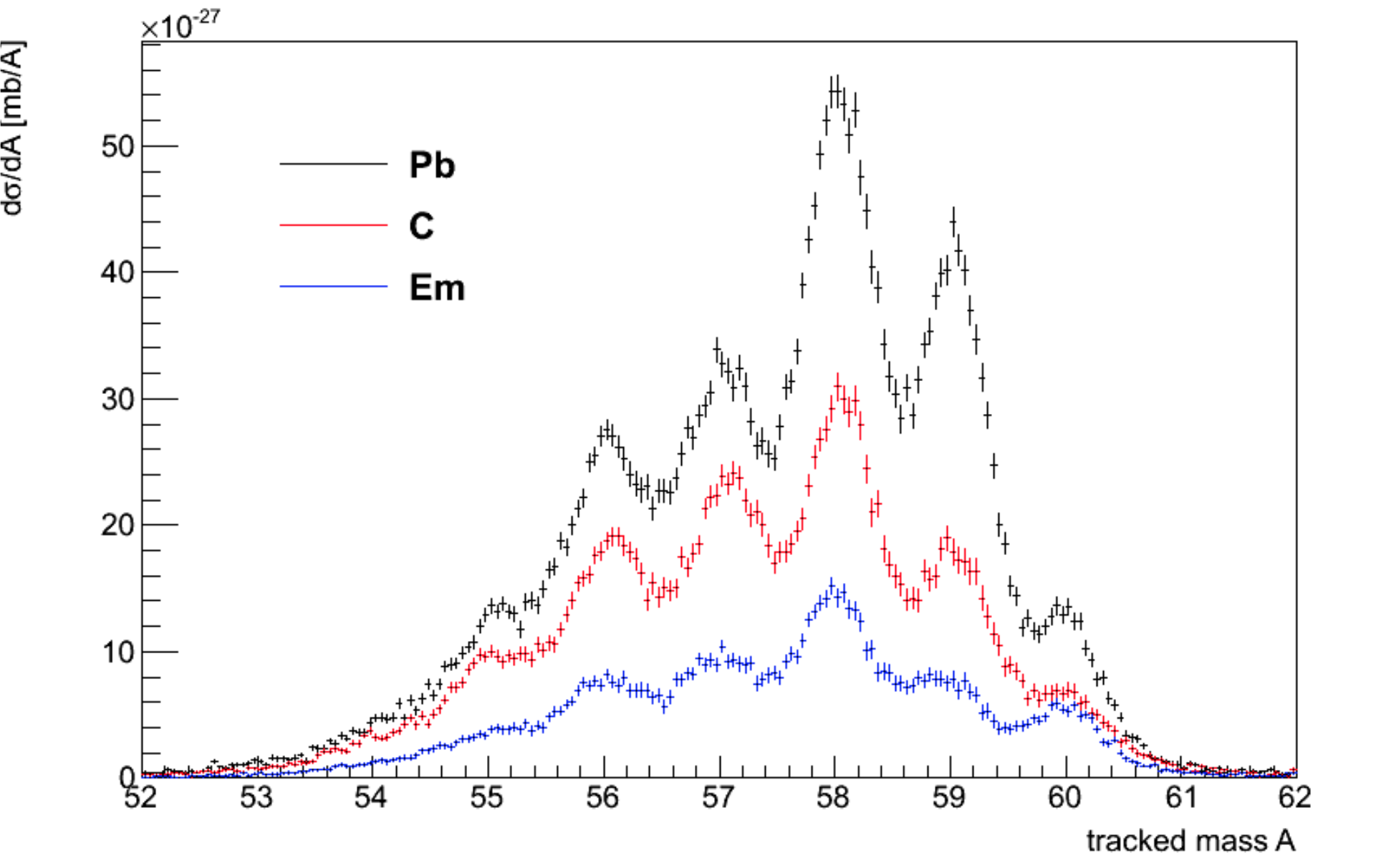}
  \includegraphics[width=.495\textwidth]{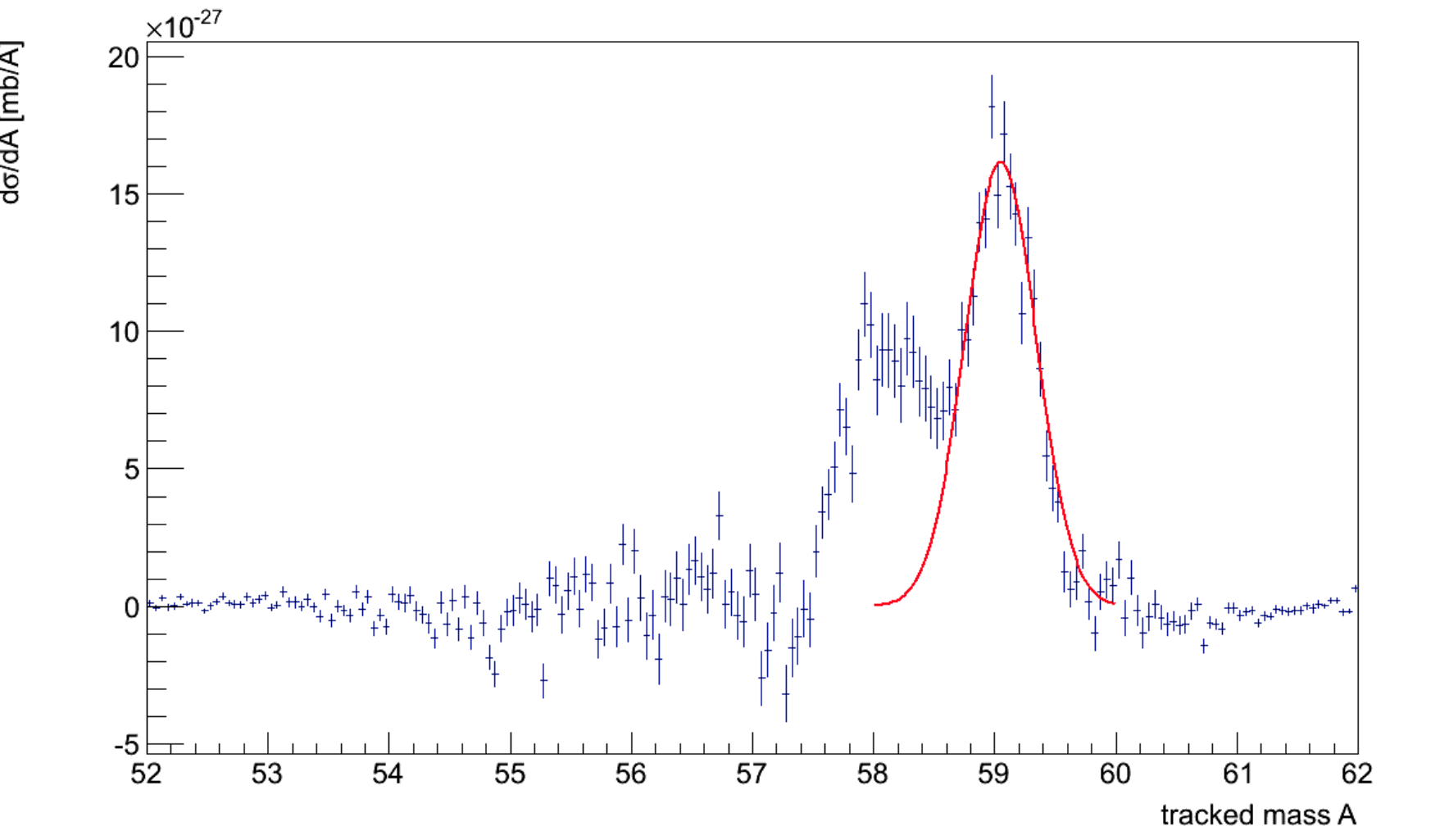}
\end{center}
  \caption{Left: Mass identification of different ion isotopes after requiring an incoming $^{60}$Fe ions a a neutron detected by LAND. Right: Spectra shown on the left subtracted such that only the mass distribution of Coulomb breakup events are left. 
  \label{fraA_60fe}}
\end{figure}

All reaction products are detected and characterised in terms of charge, mass and momentum. This allows the investigation of the neutron removal of $^{60}$Fe for different experimetal settings, Figure~\ref{fraA_60fe} (left). A lead target was used to determine the Coulomb breakup cross section, while runs with carbon and no sample at all have been performed to determine different background components.
After scaling and subtracting the contribution from nuclear interaction (measured with the carbon target) as well as interaction with other components in the beam line (empty), the pure Coulomb breakup can be extracted, Figure~\ref{fraA_60fe} (right).

\begin{figure}
\begin{center}
  \includegraphics[width=.9\textwidth]{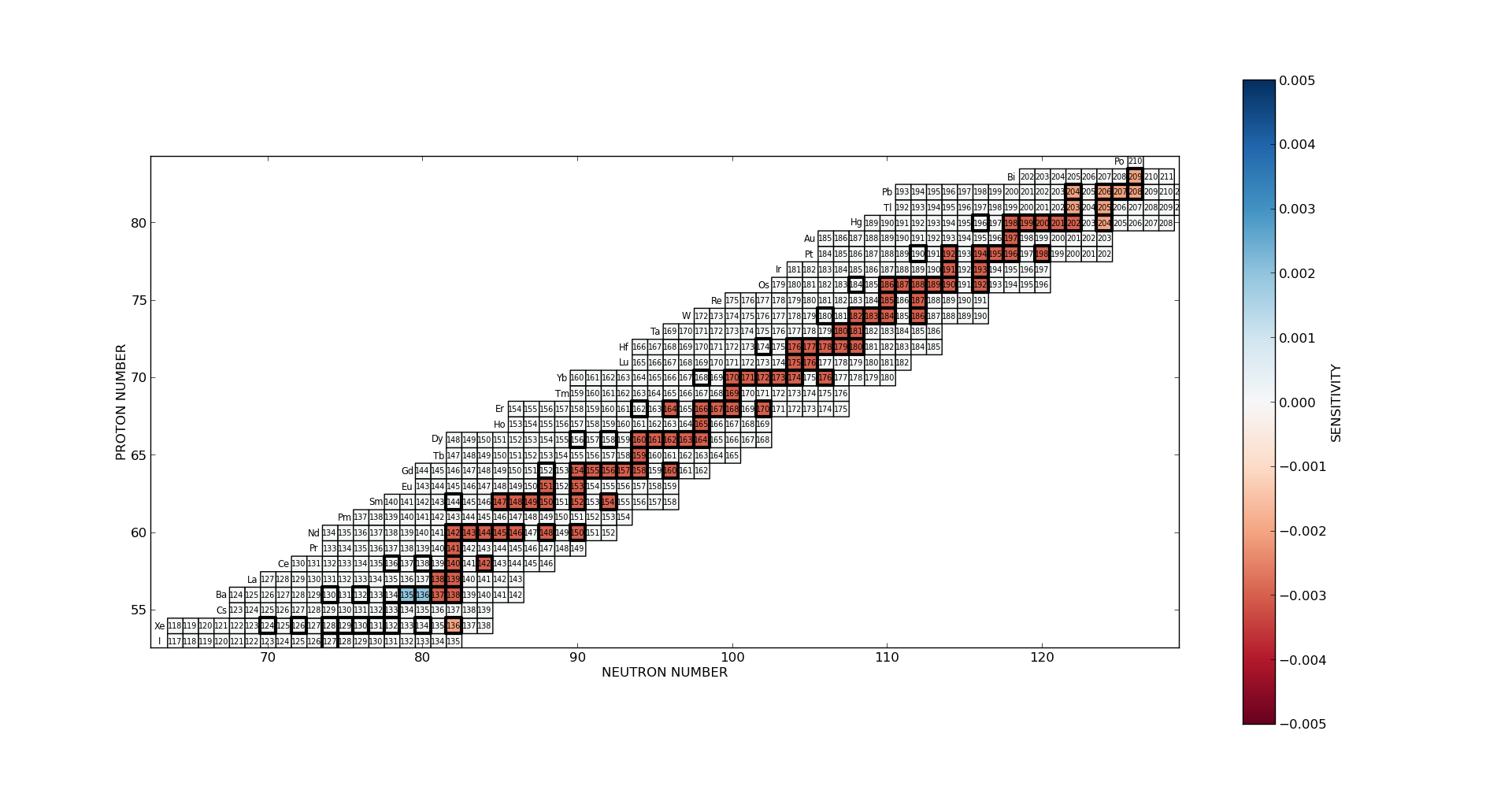}
\end{center}
  \caption{Impact of the $^{135}$I(n,$\gamma$) rate on the final abundances of the i process. This reaction rate affects most of the abundances beyond $^{135}$I and is therefore of global importance.
  \label{i_process_I135_sens}}
\end{figure}

Under certain conditions, stars may experience convective-reactive nucleosynthesis episodes. It has been shown with 
hydrodynamic simulations that neutron densities in excess of $10^{15}$~cm$^{−3}$ can be reached \cite{HPW11, GZY13}, if unprocessed, H-rich material is convectively mixed with an He-burning zone. Under such conditions, which are between the $s$ and $r$ process, the reaction flow occurs a few mass units away from the valley of stability. These conditions are sometimes referred to as the i process (intermediate process). One of the most important rates, but extremely difficult to determine, is the neutron capture on $^{135}$I, Figure~\ref{i_process_I135_sens}. The half-life time of $^{135}$I is about 6~h. Therefore the $^{135}$I(n,$\gamma$) cross section cannot be measured directly. The much improved production rates of radioactive isotopes at FAIR, however, offer the possibility to investigate the Coulomb dissociation of $^{136}$I. This reaction can then in turn be used to constrain the $^{135}$I(n,$\gamma$) rate.

\section{$^{152}$Eu - branch point in the $s$ process}\label{section_152eu}

The
laboratory based measurements of beta-decay and electron
capture rates can not directly be used in stellar
simulations. 
Electron capture can occur on excited states which are energetically
not allowed on earth \cite{LaM98}. Also beta-decays which occur from thermally
excited states cannot be measured in the laboratory. These effects
can sometimes alter the decay rates by a few orders of magnitude
\cite{TaY87,AAA04}. For the theoretical
calculations of stellar rates, Gamow-Teller strength distributions
B(GT) for low lying states are needed \cite{FFN80,LaM00,LaM03}. Charge-exchange reactions,
like the (p,n) reaction, allow access to these transitions and can
serve as input for rate calculations. In particular, there exists a
proportionality between (p,n) cross sections at low momentum
transfer (close to 0$^{\circ}$) and B(GT) values,

\begin{equation}
\frac{d\sigma^{CE}}{d\Omega}(q=0)=\hat{\sigma}_{GT}(q=0)B(GT),
\end{equation}

where $\hat{\sigma}_{GT}(q=0)$ is the unit cross section for GT
transitions at q=0. \cite{TGC87}. In order to access GT
distributions for unstable nuclei experiments have to be carried out
in inverse kinematics with radioactive ion beams. This requires the
detection of low-energy neutrons at large angles relative to the incoming
beam.

\begin{figure}
\begin{center}
  \includegraphics[width=.995\textwidth]{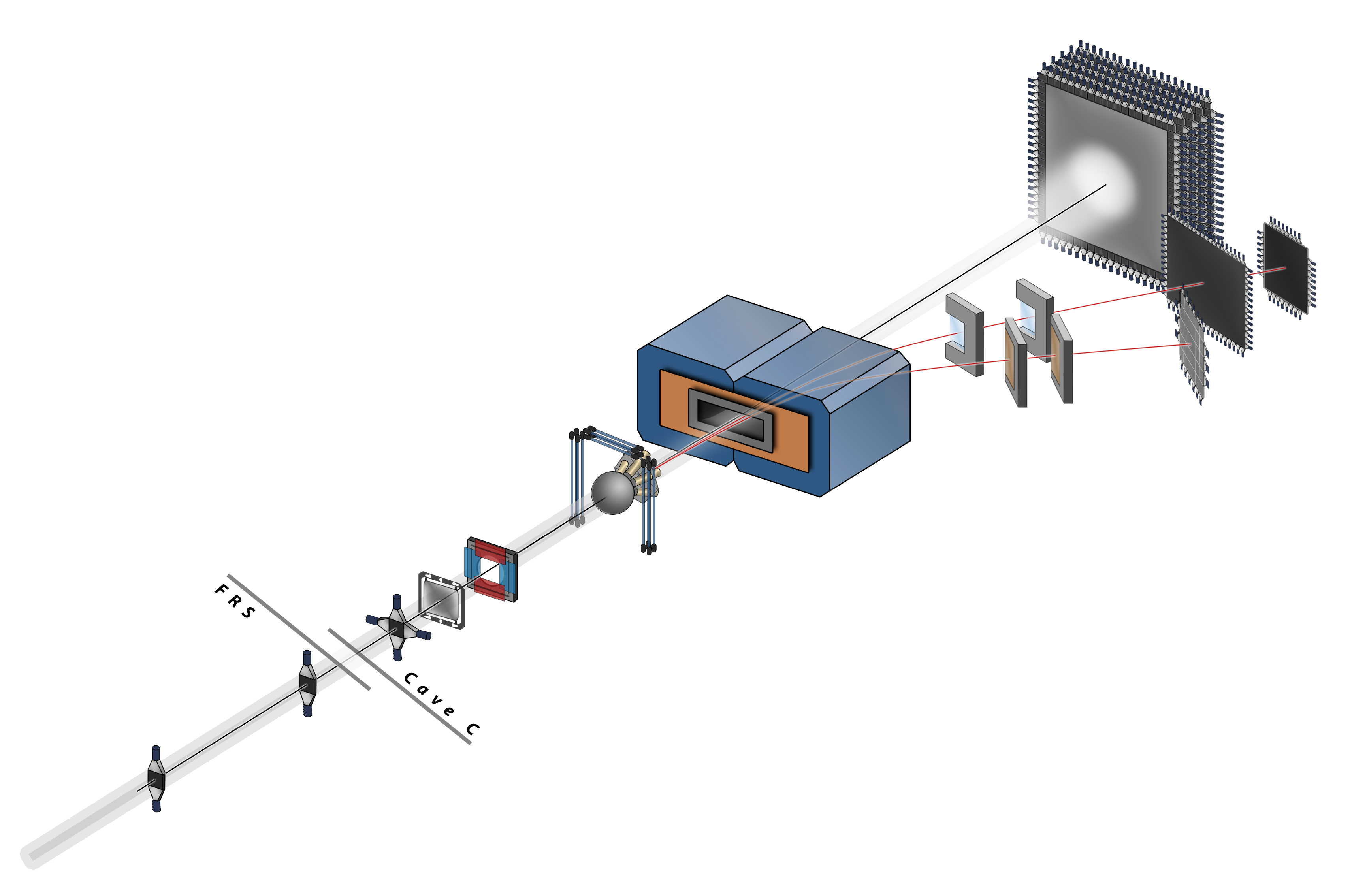}
\end{center}
  \caption{The R$^3$B setup at GSI optimized for charge-exchange reactions with low-energy neutrons in the exit channel. The LENA detector, optimized for the detection of low-energy neutrons emitted at high angles can be seen surrounding the target area. In forward direction, LaBr$_3$ detectors have been used to detect the decay of excited states.
  \label{LAND-setup1c-LENA_ohne_Silis}}
\end{figure}

An astrophysically interesting test case, $^{152}$Sm(p,n), has been investigated at GSI in inverse kinematics. In the case of inverse kinematics, all information
about the scattering angle in the center of mass and the excitation of the product 
nucleus can be determined from the energy and emission angle of the neutron in the laboratory 
reference system. Therefore a new detector for low-energy 
neutrons (LENA) has been developed \cite{LAC11} and was  used at the LAND/R3B setup, Figure~\ref{LAND-setup1c-LENA_ohne_Silis}.
The analysis of this experiment is currently ongoing and first results are very promising.

\section{Light elements in the $r$ process}\label{section_b}
\begin{figure}
\begin{center}
  \includegraphics[width=.45\textwidth]{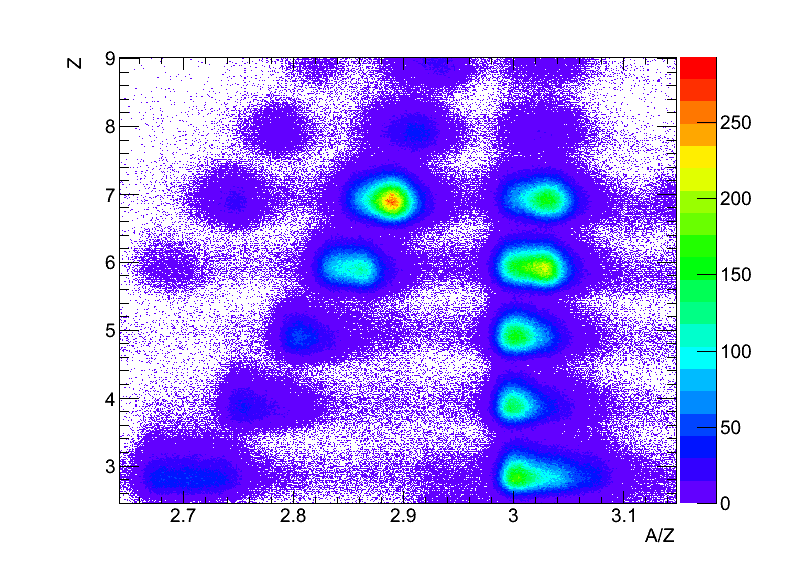}
  \includegraphics[width=.45\textwidth]{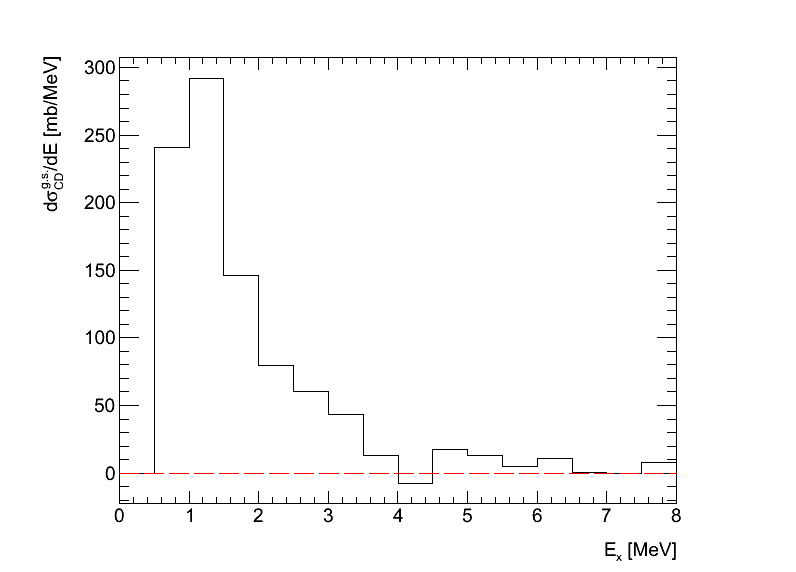}
\end{center}
  \caption{Left: Incoming identification plot. Right: Coulomb breakup cross section of $^{11}$Be at almost 500~AMeV. 
  \label{npa_proc_2013_be11}}
\end{figure}

The rapid neutron capture process
($r$ process) produces half of the elements heavier than
iron. However, the nuclear physics properties of the
involved nuclei are not well known and its astrophysical
site is not yet identified. The neutrino-driven wind
model within core-collapse supernovae are currently
one of the most promising candidates for a succesful $r$
process. These neutrino winds are thought to dissociate
all previously formed elements into protons, neutrons
and $\alpha$ particles before the seed nuclei for the $r$ process
are produced. Hence, the neutrino-driven wind model
could explain the observational fact that the abundances
of $r$ nuclei of old halo-stars are similar to our solar
$r$-process abundances \cite{TCP02}. This also indicates that the
$r$ process is a primary process and, thus, independent
of the chemical composition of the progenitor star.
Therefore, the investigation of the nuclear reactions
among light elements forming seed nuclei prior to
the $r$ process leads to a better understanding of this
process. Model calculations within a neutrino-driven
wind scenario find a crucial change in the final $r$-process
abundances by extending the nuclear reaction network
towards very light neutron-rich nuclei \cite{TSK01}. Subsequent
sensitivity studies point out the most important reactions, which include succesive (n,$\gamma$) reactions running
through the isotopic chain of the neutron-rich boron
isotopes
$^{11}$B(n,$\gamma$)$^{12}$B(n,$\gamma$)$^{13}$B(n,$\gamma$)$^{14}$B(n,$\gamma$)$^{15}$B($\beta^-$)$^{15}$C
\cite{SKM05}. Almost
all reaction rates used in these model calculations are
only known theoretically, and their uncertainties were
estimated to be at least a factor of two \cite{SKM05}. Since the
reaction rates of unstable isotopes are very difficult to
determine experimentally, neutron breakup reactions of
the neutron-rich beryllium isotopes were investigated in inverse
kinematics via Coulomb dissociation. Figure~\ref{npa_proc_2013_be11} shows the incoming identification as well as the example of Coulomb breakup of $^{11}$Be, which served as a benchmark of the measurement. The setup used for this experiment was the same as shown in Figure~\ref{r3b_setup_heavyFragments}.

\section{Break out reactions in the rp process}\label{section_30s}

\begin{figure}
\begin{center}
  \includegraphics[width=.45\textwidth]{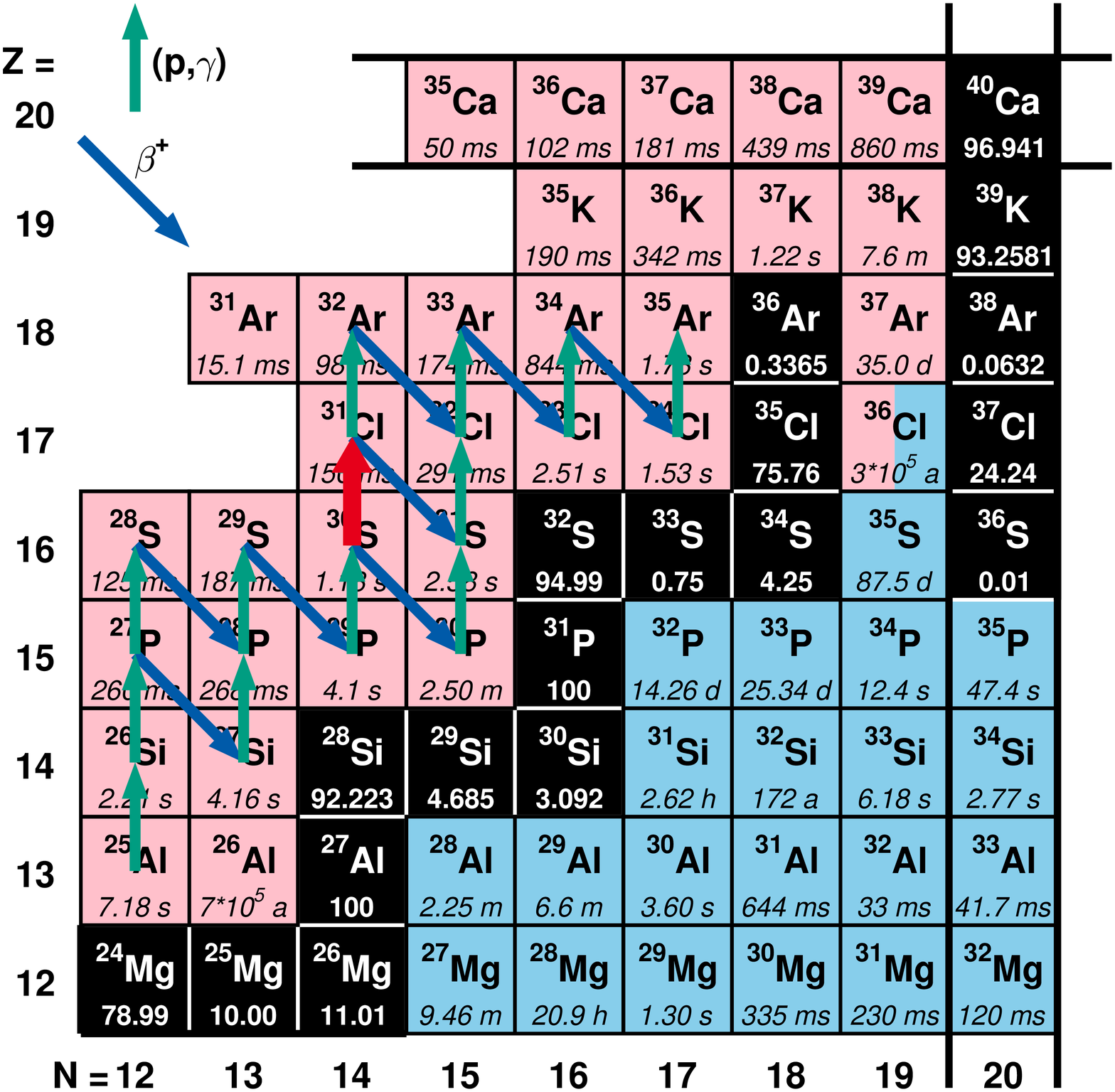}
  \includegraphics[width=.45\textwidth]{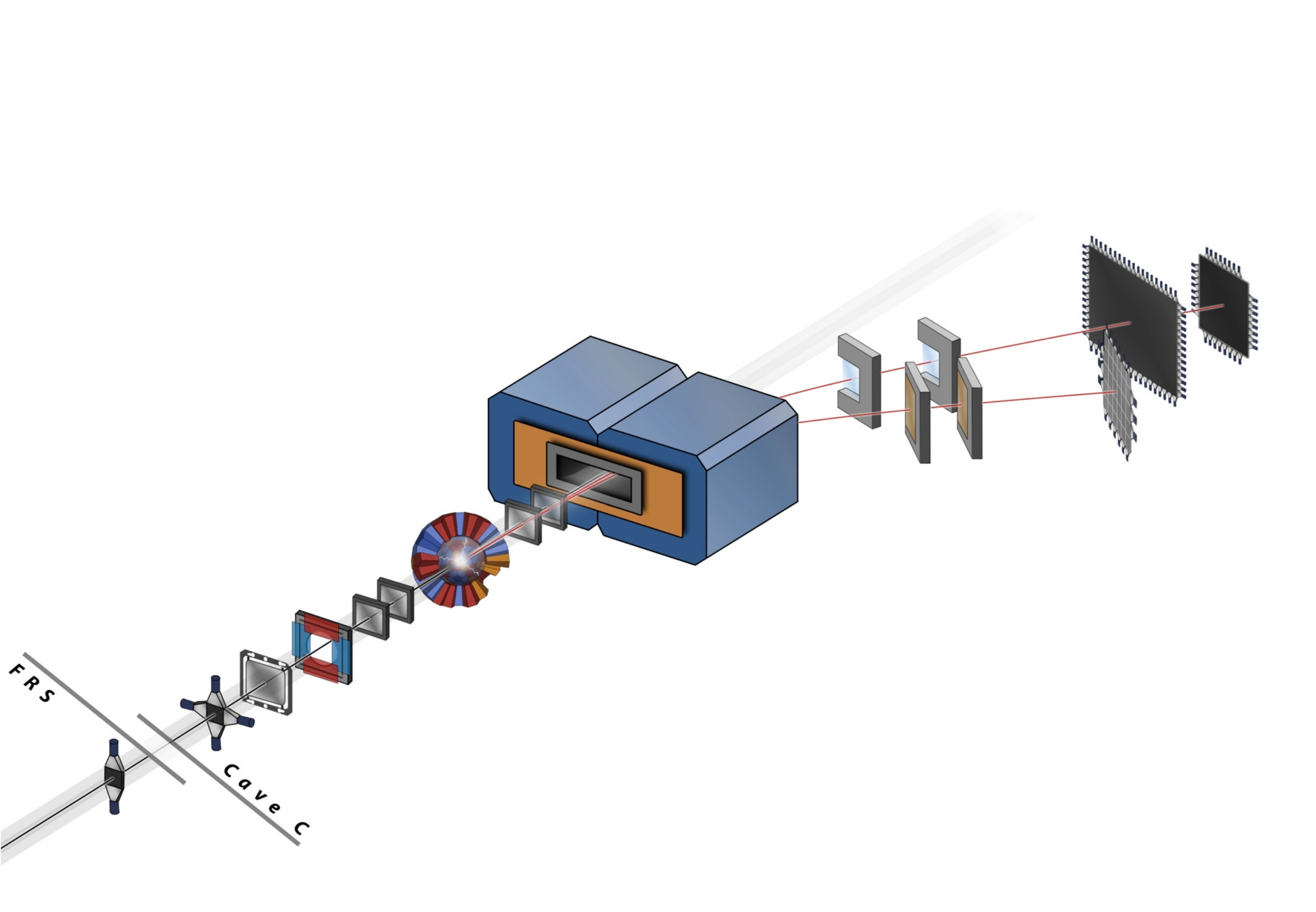}
\end{center}
  \caption{Left: The $rp$-process path at light elements. Right: The R$^3$B setup at GSI optimized for detecting protons in the exit channel. In particular the proton drift chambers are of importance. 
  \label{rp_process}}
\end{figure}

The most likely astrophysical site of X-ray
bursts are a very dense neutron stars, which accrete
H/He-rich matter from a close companion \cite{SAG98, ScR06}. While falling towards the neutron star, the
matter is heated up and a thermonuclear runaway is ignited. The exact description of
this process is dominated by the properties of a few proton-rich radioactive isotopes,
which have a low interaction probability, hence a high abundance (Figure~\ref{rp_process}, left).

\begin{figure}
\begin{center}
  \includegraphics[width=.495\textwidth]{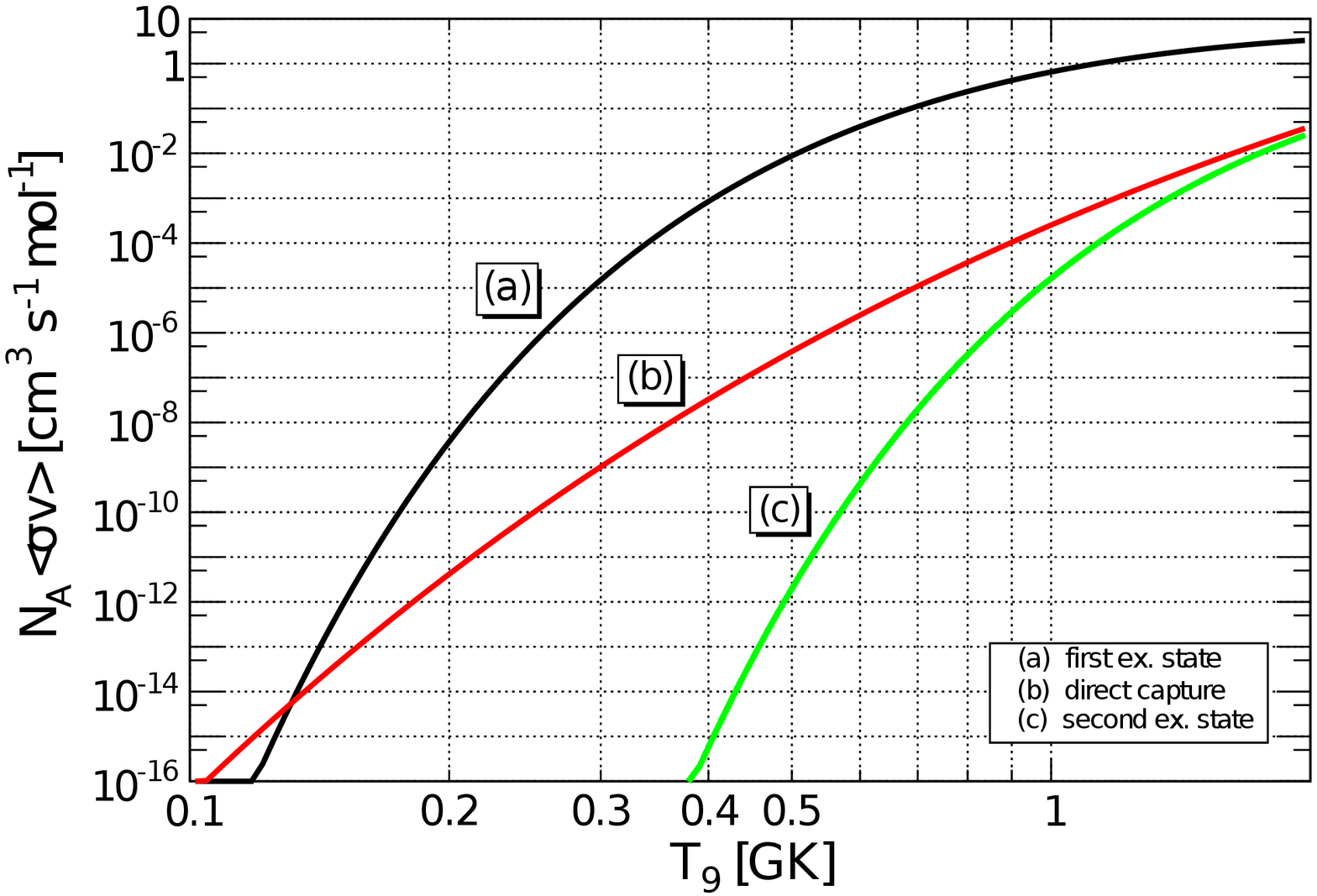}
  \includegraphics[width=.495\textwidth]{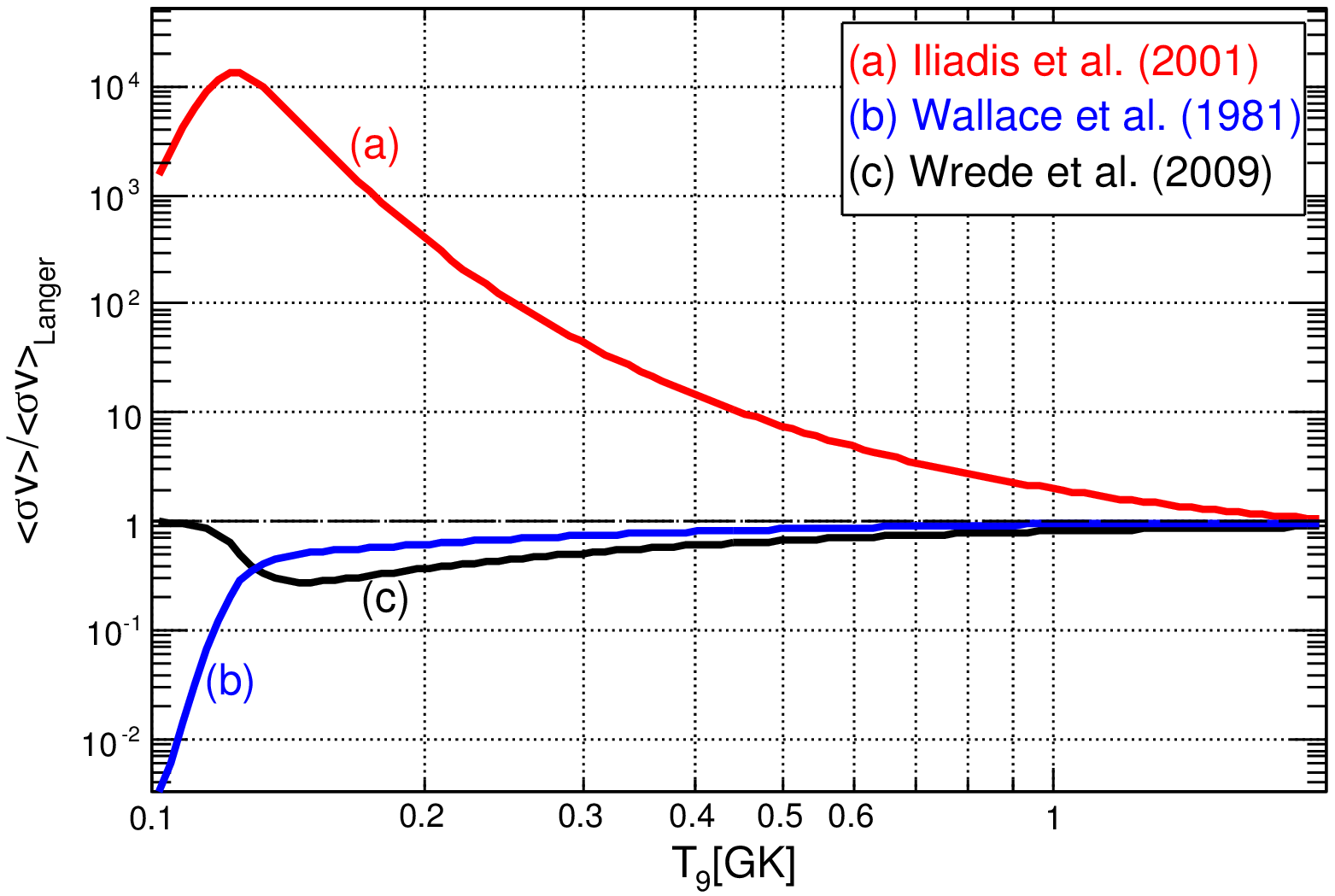}
\end{center}
  \caption{(Preliminary data) Left: The reaction rate for the important $^{30}$S(p,$\gamma$)$^{31}$Cl bottleneck reaction in the
$rp$ process. Maximum peak temperatures in the $rp$ process are typically around 2~GK. Three
contributions can be seen, from (a) the low-lying resonance, (b) the direct
capture, and (c) the second excited state in $^{31}$Cl. Right: Comparison of different previous estimations with the reaction rate derived in this work (left). a) Iliadis et al. (red) \cite{IDS01}, b) Wallace and Woosley (blue) \cite{WaW81} and c) Wrede et al (black) \cite{WCC09}. Especially in the low-temperature region,
a deviation of up to 4 orders of magnitudes is observed.
  \label{reaction_rate_30S_pg}}
\end{figure}

Therefore the short-lived, proton-rich isotopes $^{31}$Cl and $^{32}$Ar have been investigated applying the Coulomb dissociation method at the GSI. An Ar beam was accelerated to an energy of 825 AMeV
and fragmented in a beryllium target. The fragment separator was used to select the
desired isotopes with a remaining energy of 650 AMeV. They were subsequently directed
onto a $^{208}$Pb target. The measurement was performed in
inverse kinematics. All reaction products were detected and inclusive and exclusive measurements of the respective Coulomb dissociation cross sections were possible, Figure~\ref{rp_process}, right. Preliminary results for the important $^{30}$S(p,$\gamma$)$^{31}$Cl reaction and a comparison with previously known estimates are shown in Figure~\ref{reaction_rate_30S_pg}.

\section{$^{64}$Ge - a waiting point in the $\nu$p process}\label{section_64ge}
Heavy $\alpha$-nuclei are typically waiting points in the $rp$-process because of their small (p,$\gamma$) 
cross sections and the long $\beta^+$-half lives. Under certain conditions following a core collapse supernova, these waiting points can be overcome via (n,p) reactions in presence of small amount of neutrons. These neutrons stem from reactions like 
$\bar{\nu}+p \rightarrow n + \beta^+$. This process is therefore called the $\nu$p-process, \cite{TDK10}. One important waiting point is $^{64}$Ge, which implies the importance of the $^{64}$Ge(n,p)$^{64}$Ga reaction rate \cite{Fro12}. This reaction is very difficult to constrain experimentally. In combination with the planned storage rings, it would be possible to produce $^{64}$Ga beam at FAIR and store in one of the rings. In combination with a hydrogen jet target, the inverse reaction $^{64}$Ga(p,n)$^{64}$Ge could be investigated at astrophyscally interesting energies in inverse kinematics. The principle of this approach could be successfully proven with the reaction $^{96}$Ru(p,$\gamma)$, \cite{ZAB10}

\section{Summary}
Nuclear data on radioactive isotopes are extremely important for modern astrophysics. FAIR offers contributions to almost every astrophysical nucleosynthesis process. Important developments are currently ongoing while FAIR is under construction.

\ack
This project was supported by the HGF Young Investigators Project VH-NG-327, EMMI, H4F, HGS-HIRe, JINA, NAVI, DFG and ATHENA.

\section*{References}
\newcommand{\noopsort}[1]{} \newcommand{\printfirst}[2]{#1}
  \newcommand{\singleletter}[1]{#1} \newcommand{\swithchargs}[2]{#2#1}
\providecommand{\newblock}{}

\end{document}